\documentclass[%
 amsmath,amssymb,
 aps, physrev,
twocolumn,
]{revtex4-2}
\usepackage{gensymb}
\usepackage{comment}
\usepackage{graphicx}
\usepackage{bm}
\usepackage[dvipsnames]{xcolor}
\usepackage[table]{xcolor}

\definecolor{aquamarine}{rgb}{0.50,1.00,0.83}
\definecolor{mediumseagreen}{rgb}{0.24,0.70,0.44}
\definecolor{darkgreen}{rgb}{0.00,0.20,0.13}
\preprint{APS/123-QED}

\begin{document}

\title{\textbf{Density-dependent sodium-storage mechanisms in hard carbon materials}} 

\author{Alexis Front$^{a,b}$}
\email{alexis.front@aalto.fi}
\author{Tapio Ala-Nissila$^{b,c}$}
\author{Miguel A. Caro$^{a}$}
\affiliation{$^{a}$Department of Chemistry and Materials Science, Aalto University, 02150 Espoo, Finland}%
\affiliation{$^{b}$MSP Group, Department of Applied Physics, Aalto University, FI-00076 Aalto, Espoo, Finland}
\affiliation{${^c}$Interdisciplinary Centre for Mathematical Modelling and Department of Mathematical Sciences,\\ Loughborough University, Loughborough, Leicestershire LE11 3TU, United Kingdom}

\begin{abstract}
Understanding the sodium-storage mechanism in hard carbon (HC) anodes is crucial for advancing sodium-ion battery (SIB) technology. However, the intrinsic complexity of HC microstructures and their interactions with sodium remain not fully elucidated. We present a multiscale methodology that integrates grand-canonical Monte Carlo (GCMC) simulations with a machine-learning interatomic potential based on the Gaussian approximation potential (GAP) framework to investigate sodium insertion mechanisms in hard carbons with different levels of porosity, achieved by simulating structural models with densities ranging from 0.7 to 1.9 g cm$^{-3}$. Structural and thermodynamic analyses reveal the interplay between pore size and accessibility and the relative contributions of adsorption, intercalation, and pore filling to the overall storage capacity. Low-density carbons favor pore-filling, achieving extremely high capacities at near-zero voltages, whereas high-density carbons primarily store sodium through adsorption and intercalation, leading to lower but more stable capacities. Intermediate-density carbons ($1.3-1.6$ g cm$^{-3}$) provide the most balanced performance, combining moderate capacity (480 and 310 mAh g$^{-1}$), safe operating voltages, and minimal volume expansion ($<10$\%). These findings establish a direct correlation between carbon density and electrochemical behavior, providing atomic-scale insight into how hard carbon morphology governs sodium-storage. The proposed framework offers a rational design principle for optimizing HC-based SIB anodes toward high energy density and long-term cycling stability.
\end{abstract}

\maketitle

\section{Introduction}

The growing demand for renewable energy-storage solutions has exposed the limitations of conventional lithium-ion batteries (LIBs), particularly due to the scarcity of lithium in Earth’s crust \cite{Tarascon2020} and the challenges associated with its recycling. These constraints hinder the large-scale deployment of LIBs, motivating the search for alternative technologies. Sodium-ion batteries have emerged \cite{Chayambuka2020} as a promising candidate in this regard, owing to the natural abundance and low cost of sodium, as well as their favorable cycling performance \cite{Hwang2017}. Such features make sodium-ion batteries (SIBs) particularly suitable for grid-scale \cite{Hirsh2020} and large-scale energy-storage applications \cite{Slater2013}.

Graphitic carbon, the benchmark anode material for LIBs, demonstrates good compatibility with lithium ions and limited volume changes \cite{Schweidler2018}. However, its performance in SIBs is limited by the larger ionic radius of sodium, which induces significant stress and thermodynamic instability during cycling \cite{Slater2013, Yabuuchi2014}. Among alternative anode materials, hard carbon (HC) has attracted considerable attention due to its intrinsic structural stability and favorable electrochemical properties \cite{Chu2023}. Despite its widespread application, the sodium-storage mechanism in HC remains not fully understood. This complexity arises from the heterogeneous nature of HC, which comprises defects, nanoporous domains, and turbostratic graphene layers \cite{Dou2019}. These structural features strongly influence sodium-ion storage and ultimately determine the electrochemical performance of SIBs.

Sodium storage in HC proceeds through three primary mechanisms -- adsorption, intercalation, and pore filling -- which manifest as distinct features in the charge voltage profile \cite{Chen2022}. In the high-voltage region (typically above $0.2–0.3$ V vs Na$^{+}$/Na), sodium ions are predominantly adsorbed at defect sites or surface functional groups within the HC framework \cite{Surta2022}. This adsorption process leads to a rapid decrease in voltage as the limited number of high-energy sites become occupied. As the sodium content increases, the system transitions to the intercalation stage, in which sodium ions insert into the voids between graphene-like layers \cite{Huang2024}. This mechanism contributes to the sloping region of the voltage curve and is highly sensitive to interlayer spacing and the degree of structural disorder. At higher levels of sodium insertion, the pore-filling mechanism dominates. Sodium ions aggregate within larger nanopores and voids, forming clusters that give rise to the characteristic low-voltage plateau near zero voltage \cite{Stevens2000}. This stage accounts for a significant portion of the overall capacity, although it may introduce structural stress and influence long-term stability. Recently, Jian \textit{et al.} \cite{Jian2025} proposed a new interpretation of the voltage plateau origin by tailoring the pore architecture in HCs. Their study demonstrated a linear correlation between nanopore volume and the experimentally measured plateau capacity, highlighting the critical role of porosity in governing sodium-storage behavior.

Early theoretical studies of sodium storage in hard carbon primarily emphasized adsorption mechanisms, proposing that sodium ions preferentially bind to surface sites or structural defects \cite{Datta2014}. Subsequent investigations highlighted the role of enlarged interlayer spacing and structural disorder in facilitating sodium intercalation by weakening interlayer van der Waals interactions \cite{Tsai2015}. At low potentials, a pore-filling mechanism has been proposed, in which sodium ions aggregate into quasi-metallic clusters within nanopores \cite{Youn2021}. These studies form the basis of the widely accepted three-stage sodium-storage model in HC. However, most computational investigations of these mechanisms have relied on empirical interatomic potentials \cite{Surta2022, Li2022}, which provide a limited and often inaccurate description of Na–C interactions. The limited fidelity of such potentials hampers their ability to capture the diverse chemical environments present in hard carbon \cite{Caro2020}, particularly under conditions of high sodium loading. Furthermore, these models typically employ simplified structural representations that neglect essential microstructural features, including variations in mass density, pore-size distribution, defect concentration, and the presence of turbostratic graphene domains. As a result, fundamental questions remain unresolved, particularly regarding the precise role of microstructural characteristics in governing adsorption, intercalation, and filling processes, as well as the interplay between these mechanisms. Addressing these limitations requires simulation methodologies capable of accurately describing Na–C interactions while explicitly accounting for the inherent structural heterogeneity of HC.
\linebreak

In this work, we develop a machine-learning–based computational framework with density functional theory (DFT) accuracy to investigate sodium storage in hard carbon. Grand-canonical Monte Carlo (GCMC) simulations are performed using a machine-learning interatomic potential constructed within the Gaussian approximation potential (GAP) framework. This approach enables an accurate and efficient description of Na–C interactions across heterogeneous structural environments. Using this framework, we systematically examine the influence of HC porosity on sodium adsorption, intercalation, and pore filling. In particular, we identify density regimes that enable favorable sodium storage while limiting structural deformation and volume expansion. By establishing quantitative links between hard-carbon microstructure and sodium-storage behavior, this work provides methodological insights relevant to the rational design of HC anodes for sodium-ion batteries and, more broadly, to the modeling of ion storage in disordered electrode materials.
 
\section{Methodology}

To overcome the limitations of both density functional theory (DFT) and empirical interatomic potentials, there has been a recent paradigm shift to construct computationally efficient potentials by machine learning (ML) from DFT reference data. These ML methods perform high-dimensional fits to the DFT potential energy surface (PES) using a limited set of carefully selected configurations, and subsequently interpolate energies and forces for unseen structures. Unlike empirical potentials, ML models do not typically assume a fixed functional form; instead, they adapt flexibly to the training data. This reduces bias during construction but requires careful fitting and validation to prevent unphysical extrapolations. Among the different ML interatomic potential (MLIP) frameworks, GAP \cite{Bartok2010} is a kernel-based approach founded on Gaussian process regression. In this framework, the PES is regressed using fitting coefficients obtained during training in combination with kernel functions, which are evaluated on the fly. A GAP prediction is made by comparing the atomic descriptor of a current structure to a subset of structures in the database. Each comparison yields a kernel, a measure of similarity, bounded between 0 (two structures are completely different) and 1 (identical). Different descriptors of the atomic structure can be used to describe the atomic environments. The predicted local energy of an atom $i$ is then expressed for a combination of two-body (2b) and many-body (mb) descriptors as
\begin{equation}
\begin{split}
\epsilon_{i} & =e_{0} + (\delta^{(2\mathrm{b})})^{2}\sum_{s}\alpha_{s}^{(2\mathrm{b})}k^{(2\mathrm{b})}(i,s) \\
& + (\delta^{(\mathrm{mb})})^{2}\sum_{s}\alpha_{s}^{(\mathrm{mb})}k^{(\mathrm{mb})}(i,s) \\
& + \text{core},
\label{eq:1}
\end{split}
\end{equation}
where $k(i,s)$ is the kernel between the atomic environment $i$ and the different atomic environment in the sparse set (a subset of structures in the training database), the $\alpha_{s}$ are fitting coefficients obtained during training, $e_{0}$ is a constant energy per atom, $\delta$ gives the energy scale of the model, and core refers to a ``core potential''.

The choice of descriptors is central to an accurate and data-efficient PES representation. We employ a combination of 2b and mb atomic descriptors, in addition to a tabulated splined ``core potential'' to describe the interaction between two atoms when they are nearby ($\lesssim 1$~\AA{}). The 2b descriptors, constructed with a 5.5~\AA\ cutoff for C–Na and Na–Na interactions, improve the numerical stability of the GAP. To reproduce the underlying DFT PES with higher fidelity, we include many-body descriptors based on the smooth overlap of atomic positions (SOAP) formalism \cite{bartok_2013}. In this work, we use a more data-efficient implementation of the SOAP descriptor \cite{Caro2019}.

\subsection{Machine-learning interatomic potentials}

\textbf{$\Delta$-learning.} 
Modeling the potential energy surface of the Na-C system is best done by splitting its representation into two components. A high-quality GAP model for carbon is already available \cite{Muhli2021}, and it accurately captures the strong (short-range) covalent interactions and weak (long-range) van der Waals interactions in pure carbon as separate terms in the potential. By comparison, interactions between Na and C are simultaneously short-range \textit{and} weak. Therefore, rather than constructing a full C-Na binary potential from scratch, we adopt a $\Delta$-learning approach \cite{Fujikake2018}: we fit a machine-learning model only to the \emph{energy differences} induced by Na insertion into carbon matrices. The intercalation energy is defined as

\begin{equation}
\Delta E_\text{ref}(\mathrm{CNa})=E_\text{ref}(\mathrm{CNa})-E_\text{ref}(\mathrm{C}),
\end{equation}

where $E_\text{ref}(\mathrm{CNa})$ is the energy of the whole structure (sodium atoms in a carbon structure) and $E_\text{ref}(\mathrm{C})$ is the energy of the same structure after removing the sodium atoms. The subscript ``ref'' refers to the fact that these are the reference energies used to train the GAP ML model. The total system energy is then expressed as a sum of the reference carbon GAP and the CNa $\Delta$-GAP as:

\begin{equation}
E_{\mathrm{GAP}} (\mathrm{CNa})=E_{\mathrm{GAP}}(\mathrm{C})+ \Delta E_{\mathrm{GAP}}(\mathrm{CNa}).
\end{equation}

Note that because of the considerations discussed above, fitting $E_{\mathrm{GAP}}(\mathrm{C})$ and $\Delta E_{\mathrm{GAP}}(\mathrm{CNa})$ separately from $E_{\mathrm{ref}}(\mathrm{C})$ and $\Delta E_{\mathrm{ref}}(\mathrm{CNa})$, respectively, and then adding them up is a better approximation to $E_\text{ref}(\mathrm{CNa})$ than fitting $E_\text{GAP}(\mathrm{CNa})$ from $E_\text{ref}(\mathrm{CNa})$ directly. The $\Delta$ term is also able to accurately model metallic Na, which becomes relevant when intercalation takes place within nanopores larger than $\approx 1$ nm in diameter.

To improve accuracy in the short-range repulsive regime (interatomic distances below 2~\AA), tabulated pairwise core potentials for C–Na and Na–Na interactions are included. This explicit inclusion significantly improves the stability and accuracy of the GAP fit. In practice, the tabulated per-pair core interaction is removed from the DFT reference values prior to training, thereby smoothing the PES and facilitating the GAP fit. At prediction time, the tabulated interactions are reintroduced, ensuring both physical stability and fidelity to the DFT reference.\newline

\textbf{Training database and model fitting.} Initial training data were generated by randomly placing Na atoms (up to 10) in all possible C$_{60}$ isomers \cite{Sure2017}, defected graphite and nanoporous structures. The latter were generated by a melt-graphitization-quench protocol and using the GAP model \cite{Muhli2021}. This approach ensures sampling of all the relevant atomic environments that Na ions might encounter in graphitic carbon. For the extended systems, we used a simulation box of 216 atoms to recreate a vast variety of possible defects in graphite and to obtain bended graphene layers and small nanopores. The whole database was computed at the DFT level of theory with the PBE exchange-correlation functional \cite{Perdew1996} using VASP \cite{Kresse1996, Kresse1999}. Single-point DFT energies were obtained using a cutoff energy of 650~eV and one $k$ point. Four valence electrons were explicitly treated for C (2s$^{2}$ 2p$^{2}$) and seven for Na (2p$^{6}$ 3s$^{1}$). All the calculations used exactly the same convergence parameters to avoid introducing additional noise in the data.

\begin{figure}[h!]
\begin{center}
\includegraphics[scale=0.30]{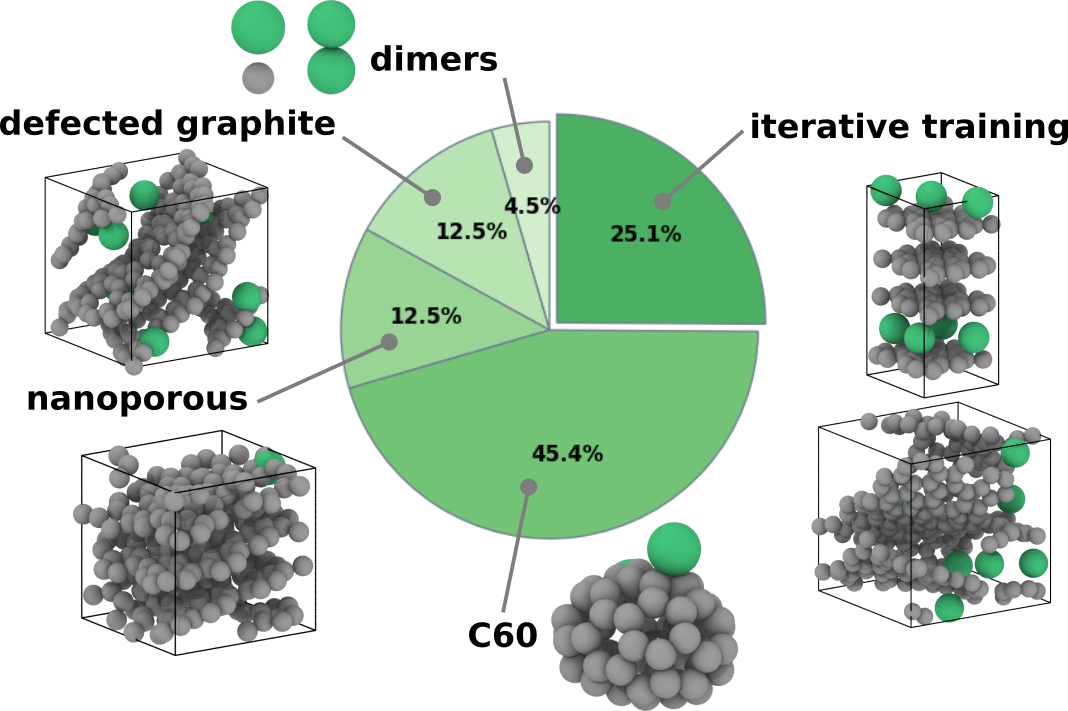}
\caption{Overview of the structures in the database. The initial database contains dimers, defected graphite, nanoporous carbon and C$_{60}$ structures. At each iteration, graphite and nanoporous carbon structures are added to the database to improve the accuracy of the model. Carbon and sodium atoms are in grey and green, respectively.}
\label{fig:1}
\end{center}
\end{figure}

To refine the model accuracy, we carried out an iterative-training approach. Starting from an initial GAP, we used short MD runs to generate 100 structures at each iteration: 25 graphite, 25 nanoporous carbon (NP-C) with a density of 1.275 g cm$^{-3}$, 25 NP-C with a density of 1.645 g cm$^{-3}$, and 25 amorphous carbon (a-C) with a density of 2.48 g cm$^{-3}$, with 1-25 sodium insertions per structure. We then ran PBE-DFT single-point calculations and added these structures to the growing training database. This procedure was repeated several times to improve the accuracy of the $\Delta$-GAP model, reaching optimal performance after the tenth cycle, with a root mean square deviation of 5 meV/at (Fig. S1). The GAP is freely available from Zenodo for further reuse \cite{Front2025_CNa}. 
The composition of the final training dataset, detailing the proportion of each structure type, is summarized in Fig. \ref{fig:1}.

All the fits were carried out with the \texttt{gap\_fit} program~\cite{Klawohn2023}, part of the \texttt{QUIP} software package~\cite{ref_quippy}. Atomic structure generation, manipulation, visualization and MD simulations were done with the Atomic Simulation Environment \cite{Hjorth2017} (\texttt{ASE}), different in-house codes, \texttt{ovito} \cite{Stukowski2010} and the \texttt{TurboGAP} program \cite{ref_turbogap}.

\subsection{Hard carbon generation}

\begin{figure*}[ht]
\begin{center}
\includegraphics[width=1.0\textwidth]{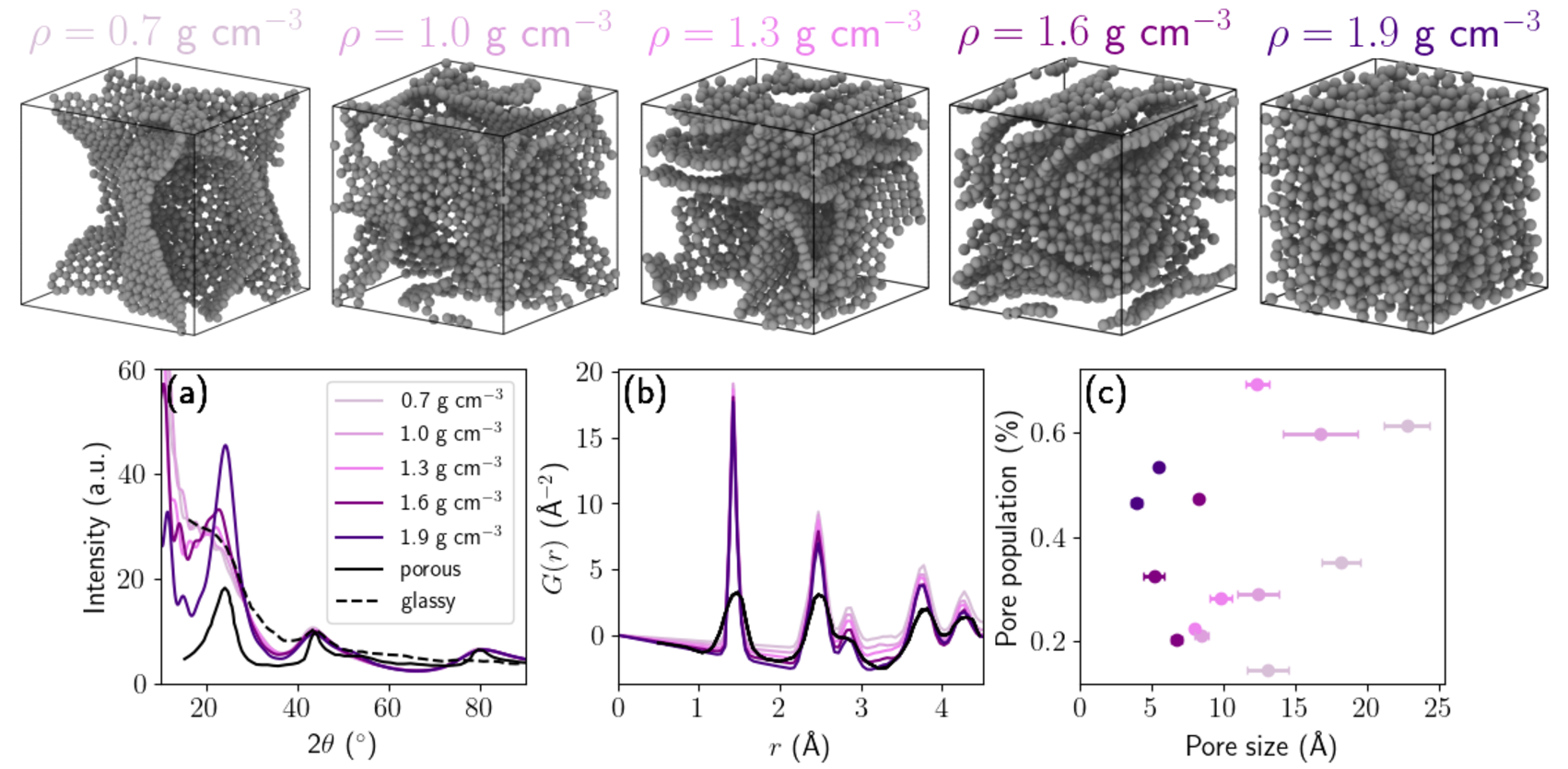}
\caption{3D snapshots of five hard carbon samples of different densities. (a) Diffraction patterns, (b) reduced partial distribution functions $G(r)$, and (c) pore-size populations. Standard deviation are represented with horizontal bars. Experimental results are extracted from \cite{Shiell2019, Shiell2021} for the HTW2500 sample (glassy) and from \cite{Huang2011} for a HC carbon sample (porous).}
\label{fig:2}
\end{center}
\end{figure*}

We used the carbon GAP of Muhli \textit{et al.} \cite{Muhli2021} which integrates Van der Waals corrections to perform molecular dynamics (MD) using the \texttt{TurboGAP} code \cite{ref_turbogap}. First, the system was equilibrated in a randomized liquid state at 9000~K for 10~ps. Next, the liquid was quenched with a linear temperature profile over 6~ps down to 3500~K. The graphitization process consist of annealing stages at 3500~K for 400~ps. Finally, the HC structures were quenched down to 300~K over 20~ps. Temperature control was achieved using the velocity-rescaling Bussi thermostat~\cite{Bussi2007} with a time constant of 100~fs. All MD simulations were performed under periodic boundary conditions with a box containing 1728 atoms, using a Verlet integration time step of 1~fs. To explore the effect of density on structural and electrochemical properties, we generated five target densities: 0.7, 1.0, 1.3, 1.6 and 1.9~g cm$^{-3}$ corresponding to cubic box lengths of 36.63, 32.52, 29.82, 27.81, 26.28~\AA \space respectively. Recognizing the stochastic nature of the melt–quench protocol and the inherent heterogeneity of HC, we generated ten independent structures per density using different random seeds. Structural properties reported in this work correspond to the mean values across these ensembles, with standard deviations capturing variability. These structures are available on Zenodo \cite{Front2025_structures}.\newline

\textbf{Structural Characterization.} 
The morphology of the quenched HC samples reveals that most of the carbon forms curved graphene-like fragments, predominantly with sp$^{2}$ bonding, as quantified in Table S1. These fragments assemble into three-dimensional networks, where edges containing sp motifs and basal planes can be interlinked via sp$^{3}$-hybridized carbon atoms. To further analyze the structural topology of the graphitic planes, we computed ring-size and bond-angle distributions \cite{Wang2022}. Here, ring size denotes the number of carbon atoms forming a closed covalent loop with the sp$^{2}$ network. These quantities serve as sensitive metrics for local order and curvature in disordered carbon networks. The analysis was performed using \texttt{matscipy} \cite{Grigorev2024} for all five HC densities. As shown in Fig. S\ref{fig:2}(a), the graphene-like layers in all samples are primarily composed of 5-, 6-, and 7-membered carbon rings. All densities display similar ring-size distributions characterized by a dominant peak at 6-membered rings, indicative of well-ordered graphitic domains, along with approximately equal populations of 5- and 7-membered rings. We also note that the small standard deviation confirms the robustness of the simulation protocol. Complementary information is provided by the bond-angle distributions. All samples feature a sharp peak at 120$^{\circ}$, a hallmark of regular hexagonal carbon rings in graphitic structures along with an inflection near 108$^{\circ}$. These pentagonal motifs, as heptagon, play an important role in curving graphene fragments in HC and glassy carbons \cite{Martin2019}. Neither the ring-size nor the bond-angle distributions show a significant dependence on density. To compare the simulated HC structures with experimental data, we computed diffraction patterns (Fig 2(a)) and reduced pair distribution functions (PDF) (Fig. 2(b)), which serve as metrics for assessing structural similarity. We chose experimental glassy sample HTW2500 reported in ref \cite{Shiell2021} and a HC carbon sample reported in ref \cite{Huang2011} as references. The reciprocal-space diffraction scattering were calculated using the Debye scattering equation as implemented in the Debyer code \cite{ref_debyer}. The reduced PDF commonly used in experiment is related to the PDF $g(r)$ most often reported in simulation work by $G(r)=4\pi\rho(g(r)-1)$, where $\rho$ is the atomic density. The HTW2500 sample exhibits three main diffraction peaks at $2\theta\sim$ 24, 44 and 80$^{\circ}$, corresponding to the graphite $\{002\}$, $\{100\}$, and $\{110\}$ reflections. For the lower-density model structures, the $\{002\}$ reflection is absent, owing to reduced and irregular layer stacking resulting from a more open and diffuse arrangement of graphene sheets. The interlayer spacing can be estimated using Bragg's law. The systematic shift toward smaller scattering angles in Fig. \ref{fig:2}(a) with decreasing density clearly indicates an increase in interlayer spacing. Among the simulated structures, only the HC sample with a density of 1.9 g cm$^{-3}$ exhibits a well-defined $\{002\}$ reflection, although the onset of this peak is already apparent for the 1.6 g cm$^{-3}$ model. From the position of this reflection, an average interlayer spacing of approximately 3.8 \AA\space can be inferred. In contrast, no significant shift is observed in the in-plane $\{100\}$ reflection, indicating an in-plane lattice parameter $a$ of around 2.43 \AA.\newline

\textbf{Pore-size distributions.} Porosity is one of the most critical properties of HC structures, as it directly influences the ability of an electrode material for ion intercalation. Therefore, characterizing porosity is a central focus of this work. We calculated the pore-size distributions using the \texttt{zeo++} code \cite{Pinheiro2013}. A three-dimensional Voronoi network \cite{Willems2012} was constructed to identify the void space within the structures, based on Voronoi decomposition as implemented in the \texttt{Voro++} library \cite{Rycroft2009}. Monte Carlo sampling was then performed to estimate the accessible pore volume and to compute the largest included sphere radius, defined as the radius of the largest sphere that can be inserted at a given point without overlapping with any of the carbon atoms in the structure. We take a probe radius of 1.16 \AA \space (ionic radius of Na$^{+}$) and 130 000 points for Monte Carlo sampling. From the resulting pore-size distributions, we extracted the mean pore radius for each density. For densities below 1.6 g cm$^{-3}$, all samples exhibit a trimodal or bimodal pore-size distribution. In contrast, at a density of 1.9 g cm$^{-3}$, only one third of the samples display a bimodal distribution, while the remaining samples show no accessible pores. Figure S2(a) presents the pore-size distributions for each structure used in the sodium insertion simulations. Porosity is defined as pore size $D$ and pore population as shown in Fig. \ref{fig:2}(c). With decreasing density, $D$ shift toward larger pore sizes, where for each density, the larger one exhibit the largest pore population. As the density increases, the PSDs also become sharper and standard deviation smaller, indicating that individual pores exhibit less size variation. A more open pore structure corresponds to larger interplanar spacing between graphitic sheets, and thus the presence of larger pores. Consequently, low-density HC may accommodate a larger number of sodium ions than high-density HC. We note that these PSDs do not explicitly account for percolation or pore connectivity, and some voids may be isolated in individual structures. Consequently, the distributions should be interpreted as indicative of typical structural trends with density rather than as exact descriptors of sodium-accessible pores.

\begin{figure*}[ht]
\begin{center}
\includegraphics[width=1.0\textwidth]{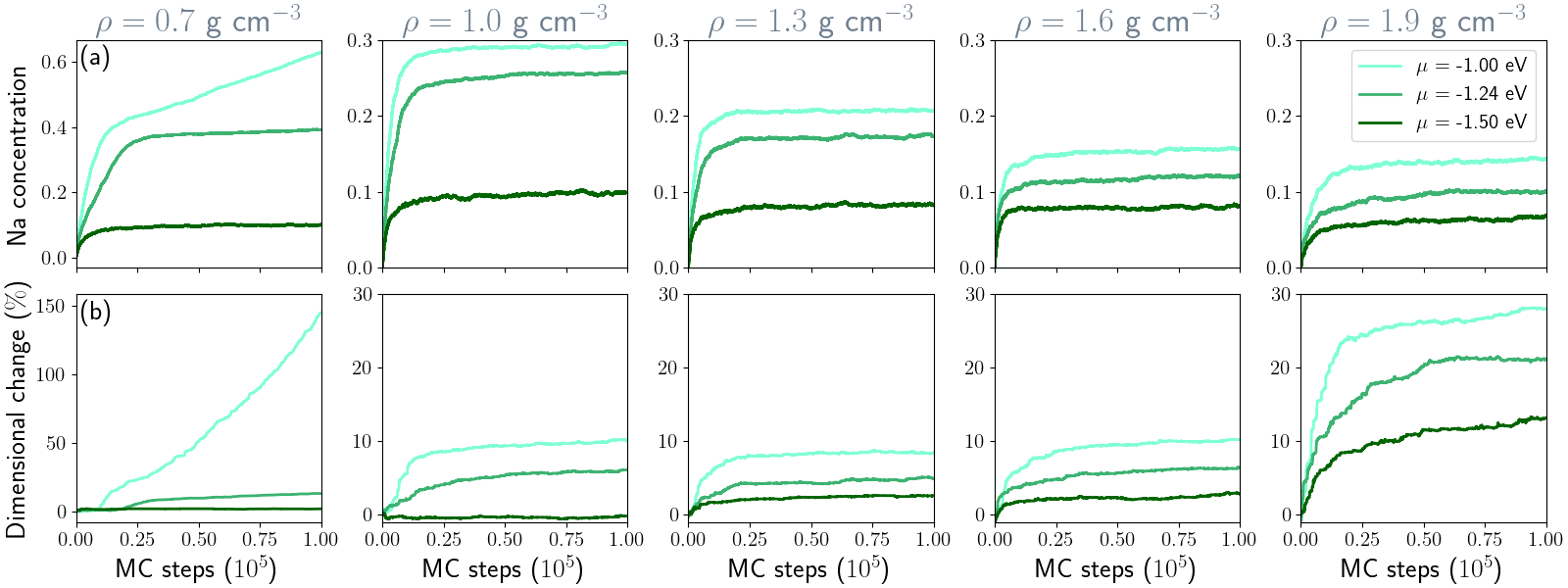}
\caption{(a) Sodium concentration and (b) dimensional change as a function of MC steps for the five hard carbon samples at 300~K. Each green line corresponds to a given chemical potential. Note the different vertical scales of the $\rho = 0.7$~g cm$^{-3}$ sample as compared to the other ones.}
\label{fig:3}
\end{center}
\end{figure*}

\subsection{Hybrid Monte Carlo/molecular dynamics atomistic simulations}

To investigate sodium-storage mechanisms in HC, we performed hybrid MC/MD simulations in the grand-canonical ensemble. In this framework, the system is coupled to a heat bath at temperature $T$ and can exchange particles with an infinite reservoir at chemical potential $\mu$. Within the Metropolis algorithm, starting from an initial pure HC structure, trial configurations are generated by one of the following moves: (i) random displacement of an atom, (ii) insertion or removal of a sodium atom at a random position, (iii) molecular dynamics steps. The inclusion of MD moves accelerates equilibration by allowing both atomic relaxations and volume adjustments. Trial configurations are then accepted or rejected according to the Metropolis acceptance probabilities for particle displacement, insertion, or removal:
\begin{equation}
P_{\mathrm{acc}}^{\mathrm{dis}}=\min[1, \exp(-\beta\Delta E)],
\end{equation}
for atomic displacements, and 
\begin{equation}
\begin{split} 
& P_{\mathrm{acc}}^{\mathrm{ins}}=\min[1, \frac{V}{(N+1)\lambda^{3}}\exp(-\beta(\Delta E-\mu))]\quad \text{and} \\
& P_{\mathrm{acc}}^{\mathrm{rem}}=\min[1, \frac{N\lambda^{3}}{V}\exp(-\beta(\Delta E+\mu))],
\end{split}
\end{equation}
for sodium insertion and removal, respectively. $\Delta E$ is the energy difference between the trial and current configurations, $\beta={1}/({k_{\mathrm{B}}T})$, $V$ is the system's volume, $N$ is the number of sodium atoms before the move, and $\lambda$ is the de Broglie thermal wavelength. A trial configuration is accepted if the corresponding acceptance probability is greater than a random number $r \in [0,1]$. Sodium insertion was simulated across the full volume of each HC structure, without explicit consideration of percolation pathways. Consequently, some sodium ions may occupy isolated voids that would be inaccessible in a real electrochemical environment. Therefore, absolute capacities obtained from these simulations may be overestimated. Nonetheless, the results allow for comparative analysis of trends in sodium insertion behavior across different structural densities.

We run MC/MD simulations for 100,000 steps for each structure for a chemical potential of $-1.00$, $-1.24$, $-1.50$~eV at 300~K. The chemical potential of $-1.24$~eV corresponds to the bcc bulk sodium cohesive energy. Figure \ref{fig:3} shows the evolution of sodium concentration and dimensional change, defined as the relative change in the volume from the hard carbon volume after inserting sodium ions. All simulations reached the equilibrium after 50,000 MC steps, except for the low density structure at the highest chemical potential ({\it i.e.}, $-1.00$~eV). The Monte Carlo simulations demonstrate a strong dependence of sodium storage on the structural density of nanoporous carbon. At low density (0.7~g cm$^{-3}$), the framework accommodates the highest sodium concentration, reflecting the abundance of adsorption sites and the ease of pore filling. As the density increases to 1.6~g cm$^{-3}$, the equilibrium concentration decreases significantly, suggesting that pore accessibility and connectivity become restricted. At the highest density investigated (1.9~g cm$^{-3}$), sodium uptake is limited to 15\%, which points to intercalation into a reduced number of energetically favorable sites rather than extended adsorption. This density-dependent trend is pronounced at $\mu=-1.00$ and $\mu=-1.24$~eV. In contrast, for $\mu=-1.50$~eV, sodium uptake is only weakly affected by density, indicating that in this regime insertion is governed primarily by the strong chemical driving force rather than by structural accessibility. 

The dimensional change of the carbon framework follows a similar trend. At low density ($\rho = 0.7$~g cm$^{-3}$) and high chemical potential, sodium insertion induces extreme swelling (up to 150 \%), directly correlated with the large Na concentration. In this case, drastic structural rearrangements occur, including the breaking of sp$^3$ bonds to accommodate volume expansion. At $\mu=-1.24$~eV, the dimensional change is much lower (13 \%), suggesting a less disruptive insertion mechanism. In the intermediate density range ($1.0–1.6$ g cm$^{-3}$) expansion remains moderate (9 and 6 \% for $-1.00$ and $-1.24$~eV respectively), pointing to a mixed regime where adsorption still contributes but is increasingly restricted by pore confinement. At high density (1.9~g cm$^{-3}$), dimensional changes strongly increase ($10–28$ \%), which can be attributed to the dominance of graphitic domains that undergo interlayer separation upon intercalation \cite{Wang2021}. This resonates with the familiar result that Na ions do not intercalate in graphite and graphite-like materials that are able to accommodate the smaller Li ions. Consistent with the Na uptake results, the dimensional change at $\mu=-1.50$~eV remains largely independent of density for structures containing pores, underscoring the overriding influence of the chemical potential in this regime, in good agreement with an experimental dimensional change around 2 \% \cite{Alptekin2020, Escher2022} for a similar sodium concentration. The final sodiated structures of each simulation are available from Zenodo \cite{Front2025_structures}.

\section{Results and discussion}

\subsection{Electrochemical properties}

\begin{figure}[h!]
\begin{center}
\includegraphics[scale=0.6]{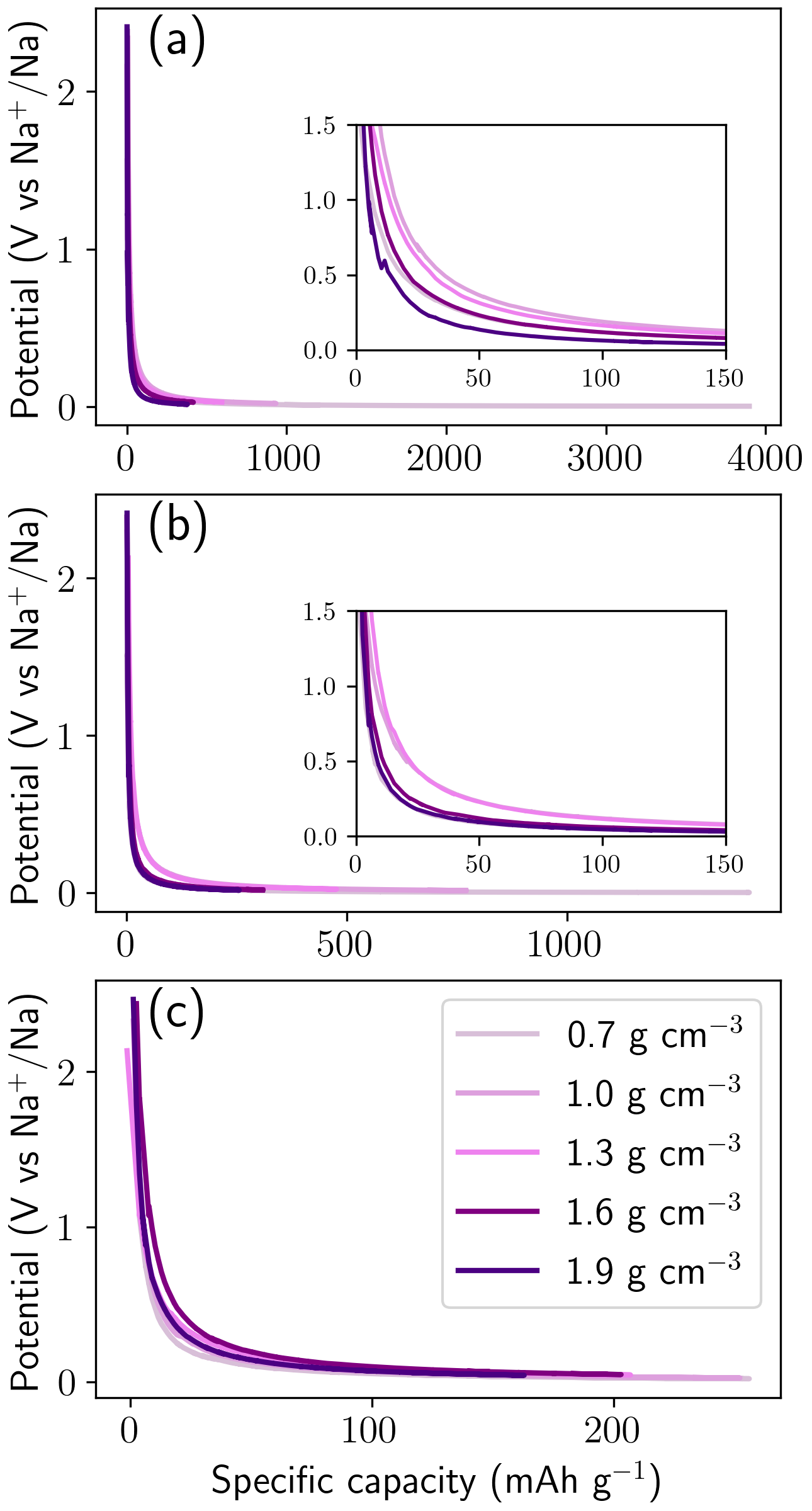}
\caption{Potential as a function of specific capacity at a given chemical potential: (a) $-1.00$~eV, (b)$-1.24$~eV, (c) $-1.50$~eV. Each line represents a different density. Insets show details of the voltage inflection.}
\label{fig:4}
\end{center}
\end{figure}

Formation energy and voltage curves are essential descriptors for understanding sodium-storage mechanisms in HC anodes. The formation energy provides direct insight into the thermodynamics of sodium insertion. By quantifying the energy change associated with placing sodium atoms into different sites of the carbon framework, it reveals the relative stability of adsorption at defects, intercalation between graphene layers, and filling of nanopores. These values help identify which storage mechanisms are energetically favored at different sodium concentrations. In contrast, the voltage curve reflects the electrochemical response of the material during sodium insertion and extraction. Derived from the slope of the formation energy with respect to sodium content, it translates atomistic energetics into an experimentally measurable quantity: the electrode potential versus Na$^{+}$/Na, expressed as
\begin{equation}
V=-\frac{1}{e}\frac{E_{\mathrm{f}}(x_{j})-E_{\mathrm{f}}(x_{i})}{x_{j}-x_{i}},
\end{equation}
where $e$ is the elementary charge of an electron, $x_{i}$ and $x_{j}$ are the sodium ions number of a state $j$ and $i$ with $x_{j}>x_{i}$, and $E_{\mathrm{f}}(x_{j})$ and $E_{\mathrm{f}}(x_{i})$ are the corresponding formation energies of these two states. Sharp drops or step-like features in the voltage curve typically correspond to transitions between different storage mechanisms, while extended plateaus indicate stable two-phase regions or clustering processes. Formation energy calculations explain why sodium ions adopt particular storage configurations, while voltage curves illustrate how these processes appear during charge–discharge cycling. This combined approach bridges atomic-level simulations with electrochemical observables, offering a comprehensive framework for evaluating and optimizing hard carbon as an anode material for sodium-ion batteries. Figure \ref{fig:4} shows the voltage curves from 0.7 to 1.9~g cm$^{-3}$ at three chemical potentials.

\begin{figure*}[ht!]
\begin{center}
\includegraphics[width=1.0\textwidth]{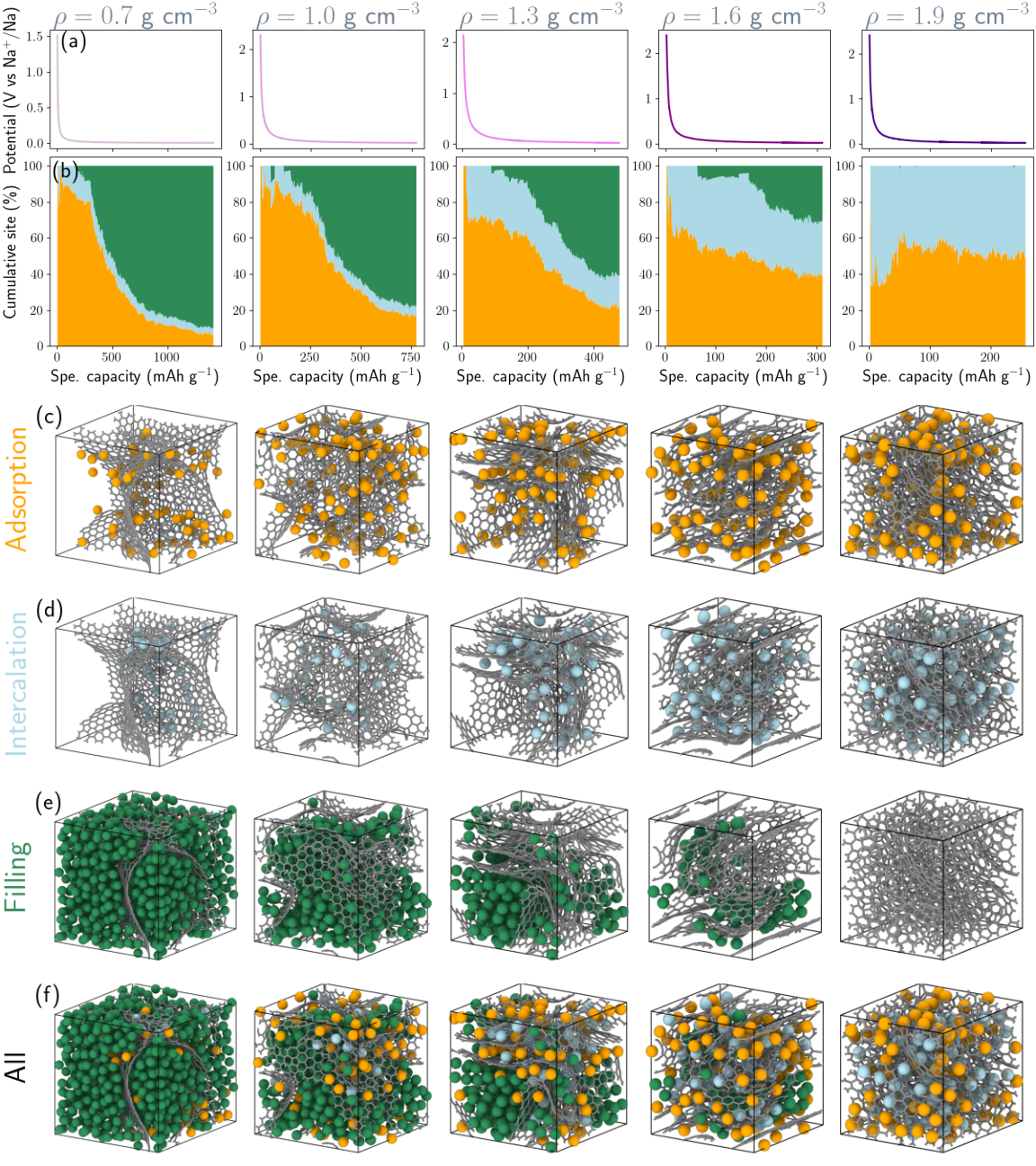}
\caption{Contributions of adsorption, intercalation, and pore filling to sodium storage in hard carbon across a density range of 0.7–1.9 g cm$^{-3}$. (a) Voltage curves showing potential versus specific capacity. (b) Cumulative storage capacity of adsorption, intercalation, and pore filling as a function of specific capacity. Snapshots in (c–e) illustrate adsorption, intercalation, and pore-filling sites, respectively. The lower panel (f) shows all three classes of storage sites together. Simulations were performed at 300~K with a chemical potential of $\mu=-1.24$~eV. All snapshots correspond to configurations at the maximum specific capacity. Color coding: orange = adsorption, blue = intercalation, green = pore-filling, and grey = carbon.}
\label{fig:5}
\end{center}
\end{figure*}

At $\mu=-1.00$~eV, the structures exhibit high storage capacities approaching 4000~mAh g$^{-1}$ in the extreme limit (which is unphysical). However, most of this uptake occurs at voltage approaching zero, which is unfavorable for practical operation due to the risk of sodium plating \cite{Bommier2015}. As chemical potential decreases, the accessible capacity is progressively reduced, reaching $\approx 1400$~mAh g$^{-1}$ at $\mu=-1.24$~eV and $\approx 250$~mAh g$^{-1}$ at $\mu=-1.50$~eV. This trend illustrates the delicate balance between the driving force for Na insertion and the stability of the host framework. The role of density is also apparent across all values of $\mu$. Higher-density structures consistently yield slightly higher voltages at low degrees of sodiation, indicating stronger stabilization of sodium within compact host environments. In contrast, lower-density systems can accommodate significantly more sodium but at lower insertion voltages because, the larger the filled pore, the more the Na within resembles bulk metallic Na. All voltage curves exhibit a sharp initial drop within the first $50$~mAh g$^{-1}$ or so, associated with the filling of the most favorable adsorption sites, followed by a gradual decay as insertion proceeds into weaker binding environments.

To rationalize these trends, Fig. \ref{fig:5} decomposes the cumulative storage capacity into three distinct contributions: adsorption, intercalation, and pore filling, and illustrates the three corresponding classes of storage sites. To quantify each contributions, we analysed the local atomic environment. Adsorption sites refer to Na-binding sites located on the carbon surface. These surface binding sites primarily consist of six-membered rings and edge motifs, with additional contributions from five- and seven-membered rings (Fig. \ref{fig:5}(c)). Intercalation sites, in contrast, correspond to Na atoms residing in interlayer regions sandwiched between two graphene sheets (Fig. \ref{fig:5}(d)). This type of binding site closely resembles those observed in the Li–graphite intercalation mechanism \cite{Stevens2001}. Filling sites correspond to sodium atoms aggregated within nanopores or voids in the carbon matrix, where Na-Na interactions dominate over Na-C bonding (Fig. \ref{fig:5}(e)). In low-density structures ($0.7-1.0$ g cm$^{-3}$) sodium uptake is initially dominated by adsorption, but at higher capacities pore filling becomes the prevailing mechanism, accounting for more than 80 \% of the total storage. This explains the very large capacities, but also why the associated voltages collapse toward zero, as pore filling corresponds to weakly bound states. At intermediate densities ($1.3–1.6$~g cm$^{-3}$), adsorption remains dominant at the onset, but intercalation contributes substantially across the entire range, while pore filling is progressively suppressed. Finally, in dense HC (1.9~g cm$^{-3}$), pore filling is negligible and sodium storage arises almost exclusively from adsorption and intercalation which lead to a limited capacity (250~mAh g$^{-1}$). The same analysis is reported in the supplementary materials for a chemical potential of $-1.00$~eV (Figure S3-(a) and S4) and $-1.50$~eV (Figure S3-(c) and S5).

These results establish a direct connection between structural density, storage mechanism, and electrochemical response. The predominance of pore filling in low-density carbons explains the experimentally observed low-voltage plateau, whereas adsorption and intercalation dominate the higher-voltage sloping region. The balance among these three mechanisms is governed by density. Low-density frameworks maximize Na uptake but operate near the sodium plating potential, while high-density carbons provide safer voltages at the cost of capacity. Intermediate-density carbons achieve the most balanced performance, combining moderate voltage, high reversibility, and limited structural deformation.

\section{Summary and conclusions}

We have developed a comprehensive methodology to investigate sodium insertion mechanisms in hard carbon. Using machine-learning potentials coupled to grand-canonical Monte Carlo simulations, we performed realistic atomistic simulations from structure generation to the evaluation of electrochemical properties across a density range from 0.7 to 1.9 g cm$^{-3}$. This approach provides direct insight into the relationship between HC porosity, sodium storage mechanism, and electrochemical performance. Our results reveal that sodium uptake in HC proceeds through three distinct contributions: adsorption, intercalation, and pore filling, whose relative importance strongly depends on mass density. Low-density HC ($0.7-1.0$ g cm$^{-3}$) achieves high capacities dominated by pore filling but operates near the potential of metallic sodium, introducing a risk of Na plating. High-density HC (1.9 g cm$^{-3}$) confines Na storage to adsorption and intercalation sites, resulting in safer operating voltages but limited capacity. Intermediate-density HC ($1.3-1.6$ g cm$^{-3}$) achieves the best trade-off, combining balanced capacity, moderate voltage, and minimal structural deformation.
These findings provide a microscopic explanation for the experimentally observed sloping and plateau regions in voltage profiles and clarify the density-driven origin of the capacity–voltage trade-off in HC anodes. The insights gained here establish density engineering as a key design strategy for next generation sodium-ion batteries, enabling the rational optimization of pore architecture, mechanical stability, and electrochemical performance.

\acknowledgements
M.A.C. and A.F. acknowledge financial support from the European Union's M-ERA.NET program (NACAB project under grant agreement No 958174). M.A.C. also acknowledge personal support from the Research Council of Finland (project number 330488). T.A-N. and A.F. have been supported in part by the Academy of Finland through its QTF Center of Excellence program (project no. 312298) and European Union -- NextGenerationEU instrument grant no. 353298. We acknowledge computational resources from CSC (the Finnish IT Center for Science) and Aalto University's Science-IT Project. 
\nocite{*}
\bibliographystyle{unsrt}
\bibliography{biblio}

\end{document}